\documentclass[letterpaper, 10 pt, conference]{ieeeconf} 

\IEEEoverridecommandlockouts 
\usepackage{mathtools}
\usepackage{amsmath}
\usepackage{amssymb}
\usepackage{mathrsfs}
\usepackage{graphicx}
\usepackage{subfigure}
\usepackage{bm}
\usepackage{color}
\usepackage{cite}

\newtheorem{asmp}{\textbf{Assumption}}
\newtheorem{lem}{\textbf{Lemma}}

\newtheorem{thm}{\textbf{Theorem}}
\newtheorem{rmk}{\textbf{Remark}}

\allowdisplaybreaks[4]

\title{\LARGE \bf
Network Weight Estimation for Binary-Valued Observation Models
}

\author{Yu Xing, Xingkang He, Haitao Fang, Karl Henrik Johansson
\thanks{This work is supported by National Key R\&D Program of China (2016YFB0901900), the National Natural Science Foundation of China (61573345), Knut \& Alice Wallenberg foundation of Sweden, and Swedish Research Council.}
\thanks{Yu Xing and Haitao Fang are with Key Lab of Systems and Control, Academy of Mathematics and Systems Science, Chinese Academy of Sciences, Beijing 100190, and School of Mathematical Sciences, University of Chinese Academy of Sciences, Beijing 100049, P. R. China
 {\tt\small yxing@amss.ac.cn; htfang@iss.ac.cn}}%
\thanks{Xingkang He and Karl Henrik Johansson are with Division of Decision and Control Systems, School of Electrical Engineering and Computer Science, KTH Royal Institute of Technology, SE-10044 Stockholm, Sweden
 {\tt\small xingkang@kth.se; kallej@kth.se}}%
}

\begin{document}

\maketitle
\thispagestyle{empty}
\pagestyle{empty}

\begin{abstract}

This paper studies the estimation of network weights for a class of systems with binary-valued observations. In these systems only quantized observations are available for the network estimation. 
Furthermore, system states are coupled with observations, and the quantization parts are unknown inherent components, which hinder the design of inputs and quantizers. 
To fulfill the estimation, we propose a recursive algorithm based on stochastic approximation techniques. 
More precisely, to deal with the temporal dependency of observations and achieve the recursive estimation of network weights, a deterministic objective function is constructed based on the likelihood function by extending the dimension of observations and applying ergodic properties of Markov chains. 
It is shown that this function is strictly concave and has unique maximum identical to the true parameter vector. 
Finally, the strong consistency of the algorithm is established. 
Our recursive algorithm can be applied to online tasks like real-time decision-making and surveillance for networked systems. 
This work also provides a new scheme for the identification of systems with quantized observations.

\end{abstract}

\section{INTRODUCTION}\label{I}

The estimation problem of networks for dynamical systems is fundamental in diverse domains such as bioinformatics, communication, as well as social networks. For example, the knowledge of gene regulatory networks can deepen our understanding of diseases and development \cite{d2000genetic}. Besides, relationship networks among individuals contain information of group structures, which is crucial for the prediction of group behavior \cite{ravazzi2017learning}. There are various formulations for the network estimation, e.g., topological inference \cite{timme2014revealing}, latent node identification \cite{nozari2018network}, etc. This paper focuses on the first one, and we define networks as weighted graphs. 

The estimation of network weights has attracted multidisciplinary attention for the last decades. \cite{timme2014revealing} reviews methods of recovering complex networks from nonlinear dynamics. Also for nonlinear systems, \cite{sharf2018network} utilizes input design and passivity approach to solve the estimation problem. Network estimation for consensus dynamics is considered in \cite{segarra2017network}, in which the estimation problem is converted to a convex optimization one. Plenty of network estimation methods for opinion dynamics, such as DeGroot and Friedkin-Johnsen models, have also been investigated, such as compressed sensing \cite{ravazzi2017learning}, vector autoregressive processes \cite{ravazzi2018randomized}, and least square algorithms \cite{dong2018identification}. 

Most existing works concentrate on systems with continuous observations. In practical scenarios, however, agents often present discrete outputs rather than continuous ones \cite{frasca2009average, ceragioli2018consensus}. For instance, binary-valued signals may be the only information transmitted and observed in communication networks because of limited storage and bandwidth resources.
Therefore, the study of network estimation for systems with quantized observations is necessary. To tackle this challenge, we resort to identification methods for quantized output systems.

The estimation of quantized systems has developed rapidly in recent years. Based on full-rank periodic inputs, \cite{wang2010system} introduces the optimal quasi-convex combination estimator. 
\cite{guo2015asymptotically} replaces the full-rank periodic inputs assumption by general quantized inputs. 
Under conditions of sufficiently rich inputs and prior knowledge of parameters, \cite{guo2013recursive, wang2018convergence} study a recursive projection algorithm for finite impulse response (FIR) systems.
Besides, input conditions can be relaxed by designing adaptive quantizers \cite{marelli2013identification, zhao2017recursive}. 
The Expectation Maximization (EM) algorithms are utilized to solve maximum likelihood estimation (MLE) problems for FIR systems in \cite{godoy2011identification} and for ARX systems in \cite{aguero2017based}, but they are batch algorithms. Finally, \cite{song2018recursive} investigates recursive identification of systems with binary outputs and ARMA noises by using stochastic approximation (SA) algorithms.

In this paper, we study the estimation of network weights for a class of binary-valued observation systems, which may not allow the design of inputs and quantizers. In these systems, agents present binary-valued outputs, which can be interpreted as true/false or active/inactive signals, and update their states based on these binary outputs. An example is quantized opinion dynamics \cite{ceragioli2018consensus}, in which agents display discrete opinions and update based on these quantized values. Other examples can be found in quantized consensus algorithms for engineering \cite{frasca2009average} and human face-to-face interactions \cite{pan2012modeling}. This update rule implies that system states are coupled with observations that cannot be modeled as selected or i.i.d. inputs as in \cite{guo2015asymptotically, guo2013recursive, song2018recursive}.
Additionally, the quantization parts of the systems are unknown inherent components and cannot be designed like in \cite{marelli2013identification, zhao2017recursive}. 

Our contributions are summarized as follows. We formulate a dynamical model over networks with binary-valued observations. The stability of outputs and the identifiability of the model are investigated in detail. To estimate network weights for this model, a recursive algorithm based on SA techniques \cite{chen2002stochastic} is proposed. More precisely, to deal with the temporal dependency of observations and achieve the recursive estimation of network weights, a deterministic objective function is constructed based on the likelihood function by extending the dimension of observations and applying ergodic properties of Markov chains. It is shown that this function is strictly concave and has unique maximum identical to the true parameter vector. Finally, the strong consistency of the proposed algorithm is established. Unlike batch algorithms solving MLE problems in \cite{godoy2011identification,aguero2017based, godoy2014identification}, our recursive algorithm can be applied to online tasks like real-time decision-making and surveillance for networked systems. This work also provides a new scheme for the identification of systems with quantized observations. 

The remainder of this paper is organized as follows. Section \ref{II} introduces some notations. We formulate the estimation problem in Section \ref{III}, and study the model and its identifiability in Section \ref{IV}. The estimation algorithm and numerical simulations are given in Section \ref{V}. Section \ref{VI} concludes the paper. 

\section{NOTATIONS}\label{II}

In this paper, we use boldfaced lower-case or Greek letters to represent column vectors. Their entries are represented by lower-case letters with corresponding subscripts, e.g., $a_i$ is the $i$-th entry of $\bm{a}$. Matrices and random vectors are written as upper-case letters such as $A$ and $X$, but we will not emphasize the meaning unless this causes ambiguity. The expectation of a random variable $X$ is denoted by $E\{X\}$.

For a matrix $A$, its entries, rows, and transpose are denoted by $a_{ij}$, $A_i$, and $A^T$, respectively. For a sequence of random vectors, say $\{X_t\}_{t \ge 0}$, $X_{k, i}$ is used to represent the $i$-th entry of $X_k$. Denote $|\bm{a}| = (|a_1|, \dots, |a_n|)^T$ and $|A| = (|a_{ij}|)$, where $|x|$ is the absolute value of real number $x$. The $n$-length all-zeros and all-ones vectors are written as $\bm{0}_n$ and $\bm{1}_n$, or simply $\bm{0}$ and $\bm{1}$. The symbol $\bm{e}_i$ denotes a unit vector with $i$-th entry being $1$. Denote $a \vee b := \max\{a, b\}$ and $a \wedge b := \min \{a, b\}$. We use $\{0, 1\}^m := \times_{i = 1}^n \{0,1\}$ to represent the Descartes product of $m$ identical binary sets $\{0, 1\}$.

For a Markov chain $\{X_t\}$ in $\Omega$, the transition probability from $x$ to $y$ is $P(x, y) = P\{X_1 = y|X_0 = x\}$, and the t-step transition probability from $x$ to $y$ is $P^t(x, y) = P\{X_t = y|X_0 = x\}$, $x, y \in \Omega$. We say that $y$ is reachable from $x$, if there exists $t \ge 1$ such that $P^t(x, y) > 0$. 

We say that $y$ is reachable from $x$, if there exists $t \ge 1$ such that $P^t(x, y) > 0$. The Markov chain is said to be irreducible, if $y$ is reachable from $x$ for all $x, y \in \Omega$. The greatest common divisor of set $\{t \ge 1: P^t(x, x) > 0\}$ is called the period of $x$, denoted by $d(x)$. The Markov chain is aperiodic if $d(x) = 1$ for all $x \in \Omega$. We call a probability distribution $\pi$ on $\Omega$ as a stationary distribution, if $\forall y \in \Omega$, $\pi(y) = \sum_{x \in \Omega} \pi(x) P(x, y)$.

\section{PROBLEM FORMULATION}\label{III}

In the sequel, suppose that the network size $n \ge 2$. The binary observation model is as follows, 

\begin{equation}\label{binary}
\begin{aligned}
Y_t &= A S_{t - 1} + D_t,\\
S_t &= \mathcal{Q}(Y_t, \bm{c}),
\end{aligned}
\end{equation}
where $t \ge 1$, $Y_t = (Y_{t, 1}, \dots, Y_{t, n})^T$, $D_t =(D_{t, 1}, \dots, D_{t,n})^T$, $S_t = (S_{t, 1}, \dots, S_{t, n})^T$ are the state vector, the disturbance, and the observation vector at time $t$ respectively. $A$ is the weight matrix, and $\bm{c} = (c_{1}, \dots, c_{n})^T$ is the unknown quantized threshold vector. $\mathcal{Q}(Y_t, \bm{c}) = (\mathbb{I}_{[Y_{t, 1} > c_1]}, \dots, \mathbb{I}_{[Y_{t, n} > c_n]})^T$ is the quantizer. Here $\mathbb{I}_A(x)$ is the indicator function such that $\mathbb{I}_A(x) = 1$ for $x \in A$ and $\mathbb{I}_A(x) = 0$ for $x \not \in A$. 

In this model, the outputs rather than states or inputs are available for individual updates. This takes place in a variety of systems such as quantized opinion dynamics \cite{ceragioli2018consensus}, human face-to-face interactions \cite{burgoon2017social, pan2012modeling}, and quantized consensus algorithms \cite{frasca2009average}.
Our main aim in this paper is to estimate the network weight matrix $A$ and the quantization threshold vector $\bm{c}$. We propose a recursive algorithm based on stochastic approximation techniques, and prove the strong consistency of the algorithm.

For weight matrix $A$, the $ij$-th entry represents the influence weight of $j$ to $i$. To cover more situations, we do not assume that the row sums of $A$ are 1. Negative weights are permitted, which represent antagonistic relationships. Without loss of generality, we assume that $|A|$ has no row with zero sum, i.e., $|A_i| \bm{1} > 0$ for all $i$, which means that every agent has certain connections with others.

The observations are only binary in this paper, but this assumption is sufficient for characterizing diverse scenarios. For example, in the voter model \cite{asavathiratham2001influence}, agents have only two choices, i.e., to vote ($1$) or not ($0$), and in human-human interactions, speaking or not can be defined as the individual outputs \cite{pan2012modeling}. 

The disturbance $D_t$ can be interpreted as the unmodeled part of the process or the summation of observation noises. We give the following standard normal assumption for it. The normal distribution assumption is not unusual for quantized systems, since it facilitates the approximation of the MLE \cite{godoy2011identification,aguero2017based, godoy2014identification}. 
\begin{asmp}\label{standardnormalasmp}
$\{D_{t, i}\}_{1 \le i \le n, t \ge 1}$ are i.i.d. standard normal random variables. 
\end{asmp}

\section{THE MODEL AND THE IDENTIFIABILITY}\label{IV}

\subsection{STOCHASTIC STABILITY}

This section investigates the stability of observations and the identifiability of the model in detail.

As in \eqref{binary}, the observation sequence $\{S_t\}$ is actually a Markov chain with finite states. The existence of stationary distributions is a significant aspect of stochastic stability of Markov chains \cite{meyn2012markov}, and Assumption \ref{standardnormalasmp} guarantees stability for observations of our model, as the following shows. 

\begin{thm}\label{MC}
Suppose that Assumption \ref{standardnormalasmp} holds. The Markov chain $\{S_t\}_{t \ge 0}$ defined by \eqref{binary} is irreducible and aperiodic, and hence converges in distribution to the unique stationary distribution positive on $\{0, 1\}^n$ from any initial distribution.
\end{thm}

Define $\tilde{S}_t := (S_t^T~ S_{t - 1}^T)^T$, $t \ge 1$. This chain is critical for our estimation. Note that $\{\tilde{S}_t\}_{t \ge 1}$ taking values in $\{0, 1\}^{2n}$ is also a Markov chain. For $t \ge 1$ and $\bm{s}_{t - 1}, \bm{s}_t$, $\bm{s}_{t + 1} \in \{0, 1\}^n$,
\begin{equation}\label{tildeS}
\begin{aligned}
&\quad~ P\left\{\tilde{S}_{t + 1} = \begin{pmatrix} \bm{s}_{t + 1} \\ \bm{s}_t \end{pmatrix} \Big| \tilde{S}_{t} = \begin{pmatrix} \bm{s}_{t} \\ \bm{s}_{t - 1} \end{pmatrix} \right\} \\
&= P\{S_{t + 1} = \bm{s}_{t + 1}|S_t = \bm{s}_t\}.
\end{aligned}
\end{equation}
So $\{\tilde{S}_t\}$ is aperiodic. For states $(\bm{s}^T~\bm{u}^T)^T, (\bm{x}^T~\bm{y}^T)^T \in \{0, 1\}^{2n}$, since $\{S_t\}$ is irreducible, we have that there exists $t \ge 1$ such that $P^t(\bm{x}, \bm{u}) > 0$. Moreover from the proof of Theorem \ref{MC}, $P(\bm{u}, \bm{s}) > 0$ holds. Hence it follows from \eqref{tildeS} that 
\[
P\left\{\tilde{S}_{t + 1} = \begin{pmatrix} \bm{s} \\ \bm{u} \end{pmatrix} \Big| \tilde{S}_{t} = \begin{pmatrix} \bm{x} \\ \bm{y} \end{pmatrix} \right\} > 0,
\]
which implies that $\{\tilde{S}_t\}$ is also irreducible, and further we have the following result:

\begin{thm}\label{MC2}
Suppose that Assumption \ref{standardnormalasmp} holds. The Markov chain $\{\tilde{S}_t\}_{t \ge 1}$ converges in distribution to the unique stationary distribution positive on $\{0, 1\}^{2n}$ from any initial distribution.
\end{thm}

The next lemma illustrates the relation between $\{S_t\}$ and the stationary distribution of $\{\tilde{S}_t\}$.

\begin{lem}\label{lem1}
Suppose that Assumption \ref{standardnormalasmp} holds, and $\tilde{S}$ is subject to the stationary distribution of $\{\tilde{S}_t\}$. Then 
\begin{equation*}
P\{\bar{S} = \bar{\bm{s}} | S = \bm{s}\} = P\{S_1 = \bar{\bm{s}} | S_0 = \bm{s}\},
\end{equation*}
for all $\bar{\bm{s}}, \bm{s} \in \{0,1\}^n$, where $\bar{S}$ and $S$ are the first and last $n$ entries of $\tilde{S}$ respectively, i.e., $\tilde{S} = (\bar{S}^T~S^T)^T$.
\end{lem}

\subsection{IDENTIFIABILITY}\label{IV.B}

One of the central concerns in system identification is whether parameters of different values can determine an identical model \cite{ljung1987system}. For model \eqref{binary}, when we fix the distribution of disturbances in advance, the answer is negative by considering the result below.

\begin{thm}\label{ident}
Suppose that Assumption \ref{standardnormalasmp} holds. Then distinct parameters $(A~ \bm{c})$ correspond to distinct Markov chain $\{S_t\}$ defined by \eqref{binary}, where $(A~\bm{c})$ is the parameter matrix of dimension $n \times (n+1)$. That is to say, for two parameter matrices $(A~\bm{c})$ and $(\hat{A}~ \hat{\bm{c}})$ such that $a_{ij} \not= \hat{a}_{ij}$ or $c_i \not= \hat{c}_i$ for some $i, j \in \{1, 2, \dots, n\}$, the corresponding Markov chains $\{S_t\}$ and $\{\hat{S}_t\}$ are not the same in the sense that their transition probability matrices are not the same.
\end{thm}

If the noise assumption is relaxed to i.i.d. normal random variables with zero mean and variance $\sigma$, then the noise distribution function is $F(x) = \Phi(\frac x\sigma)$, where $\Phi(\cdot)$ is the cumulative density function (c.d.f.) of the standard normal random variable. It follows from the proof of Theorem \ref{ident} that ${c_i}/{\sigma} = {\hat{c}_i}/{\hat{\sigma}}$, ${a_{ij}}/{\sigma} = {\hat{a}_{ij}}/{\hat{\sigma}}$, for all $1 \le i, j \le n$. 
This implies that the model \eqref{binary} is unique up to constant multiples of the parameters. 
For the situation in which the quantized threshold $c_i \not= 0$ is known for $\forall i$, the model is uniquely defined. 
In general it is not true, but we can assume that $\sigma = 1$, because the proportion of network weights that each agent gives out to different agents is the only concern, and it remains the same when the weight matrix $A$ is multiplied by a diagonal matrix with nonzero diagonal entries. 

In the literature of quantized consensus and opinion dynamics \cite{frasca2009average, ceragioli2018consensus}, the influence weight matrix is assumed to be row stochastic ($A_i \bm{1} = 1$, $\forall 1 \le i \le n$, and $a_{ij} \ge 0$, $\forall 1 \le i, j \le n$) or absolutely row stochastic ($|A_i| \bm{1} = 1$, $\forall 1 \le i \le n$). Our model can in fact capture this assumption. It is because, denoting $B = \text{diag}(a^1, \dots, a^n)$ as the diagonal matrix with diagonal entries $a_1, \dots, a_n$ with $a^i = |A_i| \bm{1}$, \eqref{binary} can be written as
\begin{equation}\label{rmk1}
\begin{aligned}
\tilde{Y}_t &= \tilde{A} S_{t - 1} + \tilde{D}_t,\\
S_{t} &= \tilde{\mathcal{Q}}(\tilde{Y}_t),
\end{aligned}
\end{equation}
where $\tilde{Y}_t = B^{-1} Y_t$, $\tilde{A} = B^{-1} A$, $\tilde{D}_t = B^{-1} D_t$, and $\tilde{\mathcal{Q}}(\tilde{Y}_t) = (\mathbb{I}_{[\tilde{Y}_{t, 1} > \tilde{c}_1]}, \dots, \mathbb{I}_{[\tilde{Y}_{t, n} > \tilde{c}_n]})^T$. Here $\tilde{c}_i = (a^i)^{-1} c_i$, and $B^{-1}$ exists since $|A_i| \bm{1} > 0$. So $\tilde{A}$ is absolutely row stochastic in \eqref{rmk1}, and $\tilde{D}_{t, i}$, $1 \le i \le n$, become heterogeneous Gaussian noises with different variances. Under this condition, the identifiability still holds.

\section{THE IDENTIFICATION ALGORITHM}\label{V}

\subsection{THE OBJECTIVE FUNCTION AND ITS PROPERTY}

Our goal is to estimate parameters $\theta := \text{vec}\big\{(A~\bm{c})\big\}$, where $(A~\bm{c})$ is a matrix of dimension $n \times (n+1)$, and $\text{vec}\{\cdot\}$ operator generates a vector from a matrix by stacking the transpose of its rows on one another. Denote $\theta^{(i)} = (A_i~c_i)^T$. To avoid ambiguity, $\theta^* := \text{vec}\big\{(A^*~\bm{c}^*)\big\} = (((\theta^*)^{(1)})^T, \dots, (\theta^*)^{(n)})^T)^T$ is used to represent the true parameters. Given observation data $\{\bm{s}^t, 0 \le t \le T\}$, the log maximum likelihood function is
\begin{align}\nonumber
&\quad~ l(n ; \theta) \\\nonumber
&= \log P\{S_t = \bm{s}^t, 0 \le t \le T\} \\\nonumber
&= \log \prod\limits_{1 \le t \le T} P\{S_t = \bm{s}^t | S_{t - 1} = \bm{s}^{t - 1}\} P\{S_0 = \bm{s}^0\} \\\nonumber
&= \log P\{S_0 = \bm{s}^0\} + \sum\limits_{1 \le t \le T} \log P\{S_t = \bm{s}^t | S_{t - 1} = \bm{s}^{t - 1}\}\\\label{gii}
&= \log P\{S_0 = \bm{s}^0\} + \sum\limits_{1 \le t \le T} \sum\limits_{1 \le i \le n} \log g_i(\tilde{\bm{s}}^t|\theta^{(i)}),
\end{align}
where $g_i(\tilde{\bm{x}}|\theta^{(i)}) := (1 - \Phi(c_i - A_{i} \bm{x}))^{\tilde{x}_i} \Phi(c_i - A_{i} \bm{x})^{1 - \tilde{x}_i}$ with $\tilde{\bm{x}}$ in $\{0, 1\}^{2n}$ and $\bm{x}$ identical to the last $n$ entries of $\tilde{\bm{x}}$, and $\tilde{\bm{s}}^t := ((\bm{s}^t)^T~(\bm{s}^{t - 1})^T)^T$.

For fixed $\theta$, $g_i(\tilde{\bm{x}}|\theta^{(i)})$ and $\nabla g_i(\tilde{\bm{x}}|\theta^{(i)})$ are bounded since $\tilde{\bm{x}}$ takes values in $\{0, 1\}^n$. Thus, from Strong Law of Large Numbers for Markov chains (Theorem 17.1.7 in \cite{meyn2012markov}), the following hold for the chain $\{\tilde{S}_t\}$ and fixed $\theta$ a.s.:
\begin{align*}
&\quad~ \lim_{T \to \infty} \frac1T \sum\nolimits_{1 \le t \le T} \sum_{1 \le i \le n} \log g_i(\tilde{S}_t|\theta^{(i)}) \\
&= E \bigg\{\sum\nolimits_{1 \le i \le n} \log g_i(\tilde{S}|\theta^{(i)}) \bigg\},
\end{align*}
\begin{align*}
&\quad~\lim_{T \to \infty} \frac1T \sum\nolimits_{1 \le t \le T} \sum_{1 \le i \le n} \nabla_{\theta^{(i)}} \log g_i(\tilde{S}_t|\theta^{(i)}) \\
&= E \bigg\{\sum\nolimits_{1 \le i \le n} \nabla_{\theta^{(i)}} \log g_i(\tilde{S}|\theta^{(i)})\bigg\}, 
\end{align*}
where $\tilde{S}$ is subject to the stationary distribution of $\{\tilde{S}_t\}$.

Therefore, the function \[E \bigg\{\sum\nolimits_{1 \le i \le n} \log g_i(\tilde{S}|\theta^{(i)})\bigg\} = \sum\nolimits_{1 \le i \le n} E \bigg\{\log g_i(\tilde{S}|\theta^{(i)})\bigg\}\] will be used to fulfill the estimation of $\theta^*$. It has an agreeable property: 

\begin{thm}\label{mainthm1}
Under Assumption \ref{standardnormalasmp}, the true parameter vector $\theta^*$ is the unique maximum of the function $E \{\sum_{1 \le i \le n} \log g_i(\tilde{S}|\theta^{(i)})\}$, and the unique solution of the equation $\nabla_{\theta} E \{\sum_{1 \le i \le n} \log g_i(\tilde{S}|\theta^{(i)})\} = 0$, where $\tilde{S}$ is subject to the stationary distribution of $\{\tilde{S}_t\}$.
\end{thm}

\subsection{THE ESTIMATION ALGORITHM}

We use the SA algorithm to address the estimation problem for the binary observation model. For $1 \le i \le n$ and $t \ge 1$, denote
\[
K_{i}(\theta^{(i)}, \tilde{S}_{t + 1}) := \nabla_{\theta^{(i)}} \log g_i(\tilde{S}_{t + 1} | \theta^{(i)}), \]
\[
K(\theta, \tilde{S}_{t + 1}) := (K_1(\theta^{(1)}, \tilde{S}_{t + 1})^T, \dots, K_n(\theta^{(n)}, \tilde{S}_{t + 1})^T)^T,
\]
where $\theta = ((\theta^{(1)})^T, \dots, (\theta^{(n)})^T)^T$, and $g_i(\tilde{\bm{x}}|\theta^{(i)}) := (1 - \Phi(c_i - A_{i} \bm{x}))^{\tilde{x}_i} \Phi(c_i - A_{i} \bm{x})^{1 - \tilde{x}_i}$ with $\tilde{\bm{x}}$ in $\{0, 1\}^{2n}$ and $\bm{x}$ identical to the last $n$ entries of $\tilde{\bm{x}}$.

The estimation algorithm is as follows:
\begin{equation}\label{IdenAlg}
\theta_{t + 1} = \theta_t + a_t K(\theta_t, \tilde{S}_{t + 1}),
\end{equation}
where $\theta_t = ((\theta^{(1)}_t)^T,\dots,(\theta^{(n)}_t)^T)^T$ is the estimation of $\theta$ at time step $t$, and $a_t$ is the step size.

\begin{rmk}
In this algorithm, we assume that $\theta_t$ is bounded. If this assumption does not hold, one can apply the SA algorithm with expanding truncations \cite{chen2002stochastic}, in which estimate $\theta_t$ is also bounded because of truncation. It is also verified that the times of truncation is finite a.s.
\end{rmk} 

\begin{asmp}\label{at}
Let $a_t$ be the step size in \eqref{IdenAlg}, satisfying $a_t > 0$, $\sum\nolimits_{t = 1}^{\infty} a_t = \infty$, and $\sum\nolimits_{t = 1}^{\infty} a_t^2 < \infty$.
\end{asmp}

\begin{thm}\label{mainthm2}
Suppose that Assumptions \ref{standardnormalasmp} and \ref{at} hold. Then the estimates $\theta_t$ of the algorithm \eqref{IdenAlg} converge to $\theta^*$ a.s. from any fixed initial value, where $\theta^*$ is the true parameter vector.
\end{thm}

\subsection{NUMERICAL SIMULATIONS}
We use an influence weight matrix with four individuals from an empirical study \cite{friedkin1990social} to illustrate the consistency of the above algorithm. The weight matrix $\tilde{A}$ is given by
\begin{equation*}
\tilde{A} = \begin{bmatrix} .220 & .120 & .360 & .300 \\
.147 & .215 & .344 & .294 \\
0 & 0 & 1 & 0 \\
.090 & .178 & .446 & .286
\end{bmatrix}.
\end{equation*}
The noises are set to be independent Gaussian with zero mean and variance $4$, and $\tilde{c}$ is randomly selected as $\tilde{\bm{c}} = (0.13~0.28~0.08~0.24)^T$. Therefore, as previous discussion, the parameters are identical to that $\bm{c} = (0.065~0.14~0.04~0.12)^T$ and $A = \tilde{A}/2$ in our model.

We set the step size $a_t = 10 / (t + 200)$, and run the algorithm for $100$ trials. Fig. $1$ shows the strong consistency of the algorithm, illustrated by two parameters $a_{12}$ and $a_{33}$. The blue line represents one sample path, and the red line represents the true value. The gray ones are sample paths for all $100$ trials. Fig. $2$ shows the mean square error (MSE), which is defined as $\text{MSE}_k := \frac1N \sum_{i = 1}^N \|\theta_k - \theta^*\|^2$ with the number of trials $N = 100$.

\begin{figure}[t]
\centering
\subfigure{
\includegraphics[scale=0.26]{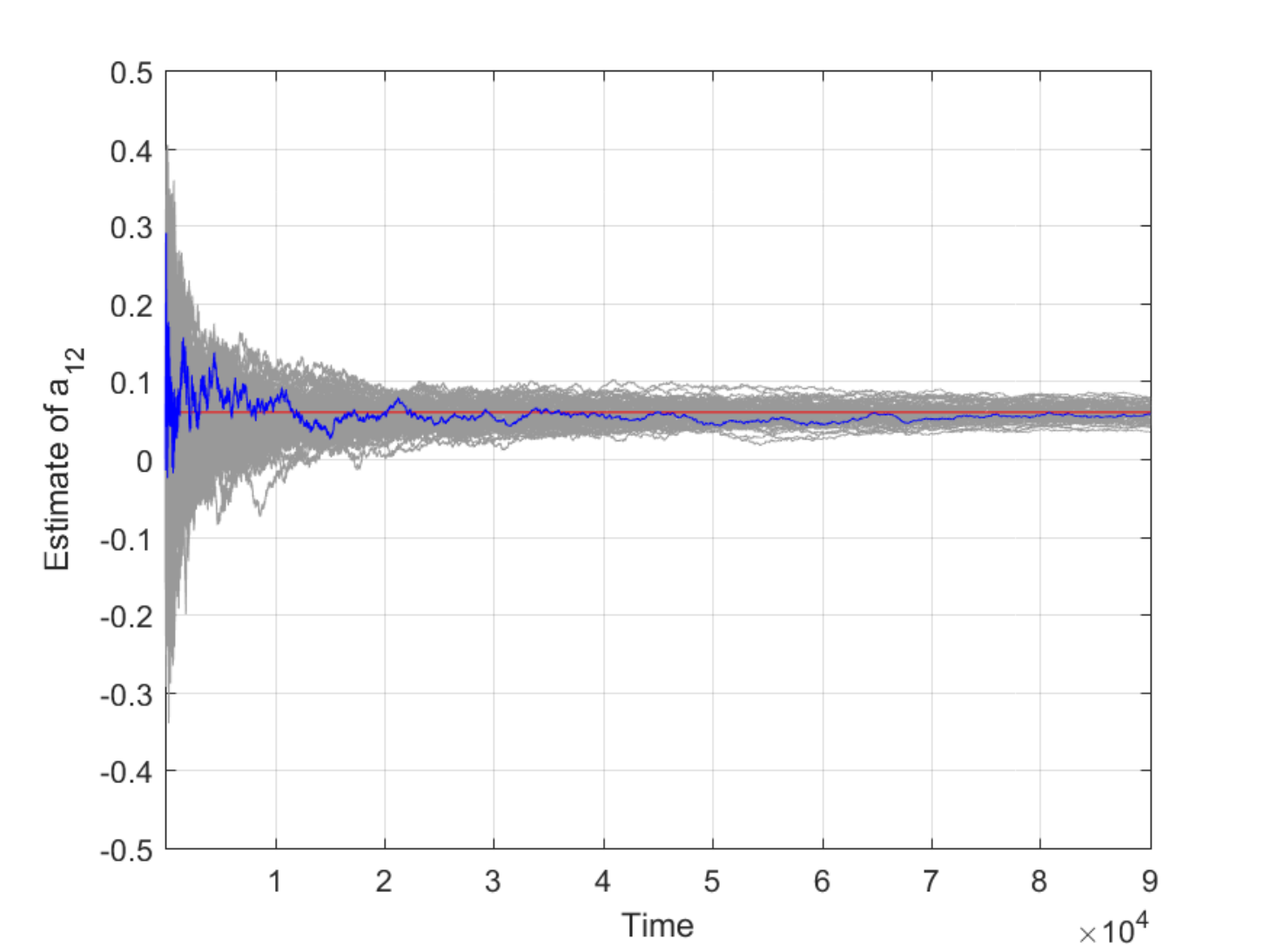}}
\subfigure{
\includegraphics[scale=0.26]{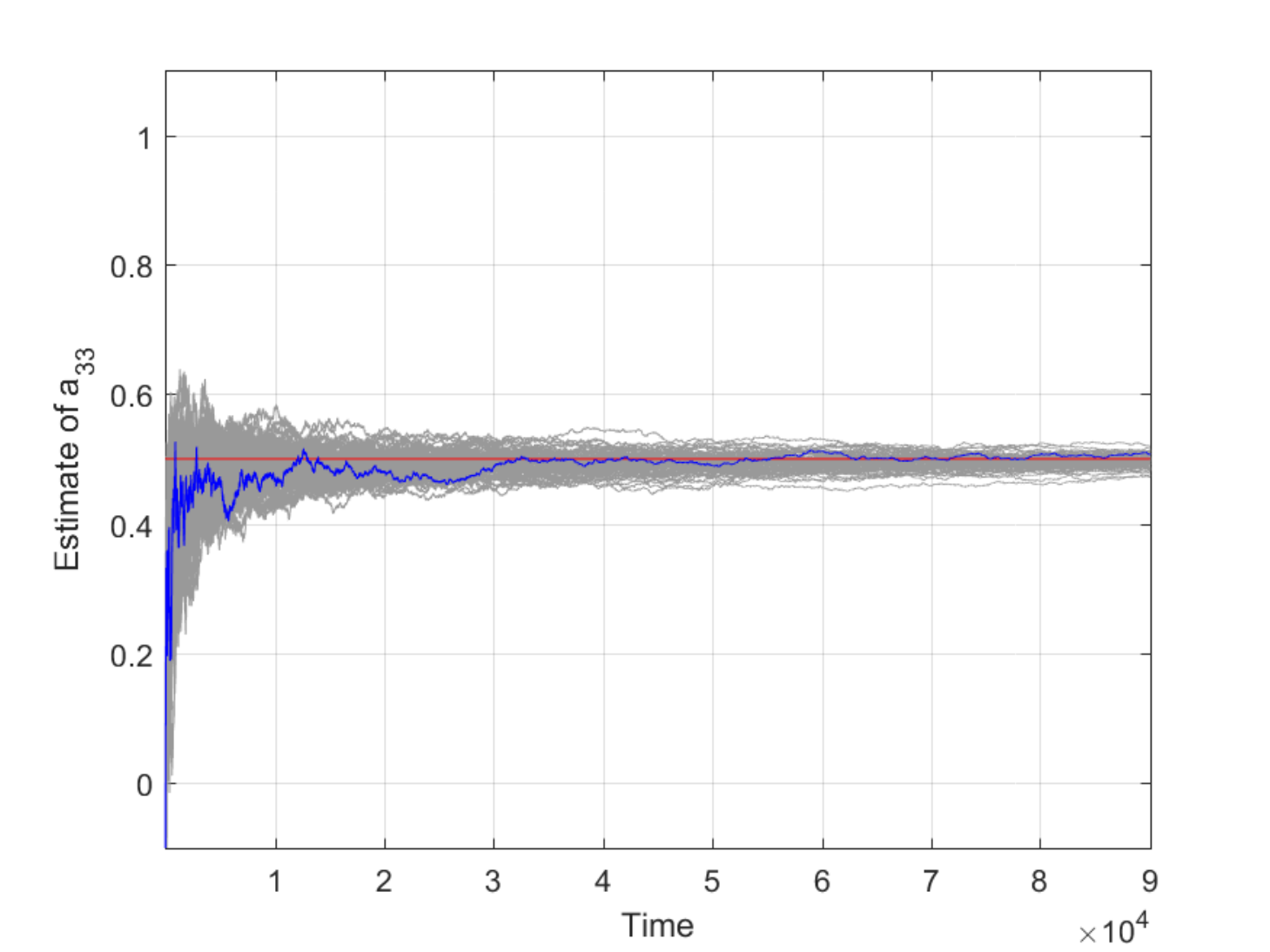}}
\caption{The strong consistency of the proposed algorithm.}
\end{figure}

\begin{figure}[t]
\centering
\includegraphics[scale=0.4]{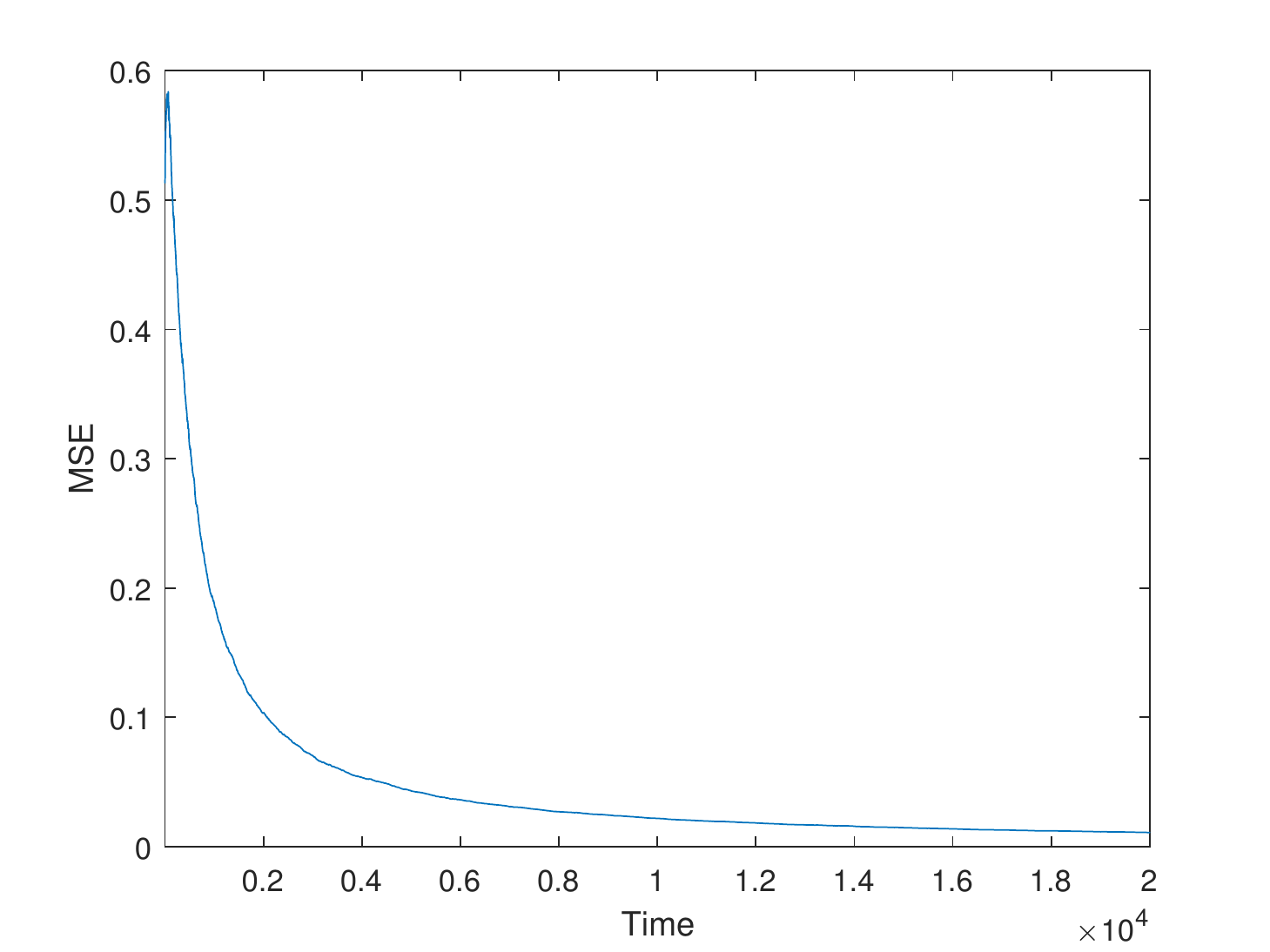}
\caption{The MSE of the proposed algorithm.}
\end{figure}

\section{CONCLUSION}\label{VI}

In this paper we study the estimation of network weights for a class of binary observation systems. These systems are distinctly different from models studied in the literature of quantized identification, because there is no room for the design of inputs and quantizers. We propose a recursive algorithm based on stochastic approximation techniques, and prove its consistency. Future work includes investigation of the convergence rate and asymptotical efficiency, generalization of the model and noise conditions, and application of the algorithm in practice.

%


\section*{APPENDIX}

\emph{\textbf{Proof of Theorem \ref{MC}}:}

Under Assumption \ref{standardnormalasmp}, the probability transition matrix can be obtained via the following way:
\begin{align}\label{gi}\nonumber
&\quad~P\{S_1 = \bm{s} | S_0 = \bm{u}\} \\\nonumber
&=P\{A_{i} S_0 + D_{1, i} > c_i, \forall i \text{ s.t } s_i =1, A_{j} S_0 + D_{1, j} \le c_j, \\\nonumber
&\qquad~ \forall j \text{ s.t. } s_j =0| S_0 = \bm{u}\}\\\nonumber
&=P\{A_{i} \bm{u} + D_{1, i} > c_i, \forall i \text{ s.t } s_i =1, A_{j} \bm{u} + D_{1, j} \le c_j, \\\nonumber
&\qquad~ \forall j \text{ s.t. } s_j =0| S_0 = \bm{u}\}\\\nonumber
&=P\{A_{i} \bm{u} + D_{1, i} > c_i, \forall i \text{ s.t } s_i =1, A_{j} \bm{u} + D_{1, j} \le c_j, \\\nonumber
&\qquad~ \forall j \text{ s.t. } s_j =0\}\\
&= \prod_{1 \le i \le n} (1 - F(c_i - A_{i} \bm{u}))^{s_i} F(c_i - A_{i} \bm{u})^{1 - s_i} > 0,
\end{align}
for all $\bm{s}, \bm{u} \in \{0,1\}^n$, $1 \le i \le n$. Therefore, the transition matrix of $\{S_t\}$ is irreducible and aperiodic, and the conclusion holds by Corollary 1.17 and Theorem 4.9 in \cite{levin2017markov}.
\hfill$\Box$

\emph{\textbf{Proof of Lemma \ref{lem1}}:}

Let $\tilde{P}$ be the transition probability matrix of $\{\tilde{S}_t\}$. From the definition of stationary distribution, we have that
\begin{equation*}
P\{\tilde{S} = (\bar{\bm{s}}^T~\bm{s}^T)^T\} = \sum_{\tilde{\bm{s}} \in \{0, 1\}^{2n}} P\{\tilde{S} = \tilde{\bm{s}}\} \tilde{P} (\tilde{\bm{s}}, (\bar{\bm{s}}^T~\bm{s}^T)^T).
\end{equation*}
Define \[\mathscr{S}_1 := \{\tilde{\bm{s}} \in \{0, 1\}^{2n} : \text{the first } n \text{ entries of } \tilde{\bm{s}} \text{ are } \bm{s}\},\] and it follows from the definition of $\{\tilde{S}_t\}$ that $P(\tilde{\bm{s}}, (\bar{\bm{s}}^T~\bm{s}^T)^T) = 0$ for $\tilde{\bm{s}} \not \in \mathscr{S}_1$. Hence, 
\begin{equation}\label{lem1eq1}
P\{\tilde{S} = (\bar{\bm{s}}^T~\bm{s}^T)^T\} = \sum_{\tilde{\bm{s}} \in \mathscr{S}_1} P\{\tilde{S} = \tilde{\bm{s}}\} \tilde{P} (\tilde{\bm{s}}, (\bar{\bm{s}}^T~\bm{s}^T)^T).
\end{equation}
Similarly, we have that
\begin{align}\nonumber
P\{S = \bm{s}\} &= \sum_{\tilde{\bm{s}}^2 \in \mathscr{S}_2} P\{\tilde{S} = \tilde{\bm{s}}^2\}\\\label{lem1eq2}
&= \sum_{\tilde{\bm{s}}^1 \in \mathscr{S}_1} \sum_{\tilde{\bm{s}}^2 \in \mathscr{S}_2} P\{\tilde{S} = \tilde{\bm{s}}^1\} \tilde{P} (\tilde{\bm{s}}^1, \tilde{\bm{s}}^2),
\end{align}
where \[\mathscr{S}_2 := \{\tilde{\bm{s}} \in \{0, 1\}^{2n} : \text{the last } n \text{ entries of } \tilde{\bm{s}} \text{ are } \bm{s}\}.\] Combining \eqref{tildeS} \eqref{lem1eq1} and \eqref{lem1eq2},
\begin{align*}
&P\{\tilde{S} = (\bar{\bm{s}}^T~\bm{s}^T)^T\} = \sum_{\tilde{\bm{s}} \in \mathscr{S}_1} P\{\tilde{S} = \tilde{\bm{s}}\} P (\bm{s}, \bar{\bm{s}}),\\
&P\{S = \bm{s}\} = \sum_{\tilde{\bm{s}}^1 \in \mathscr{S}_1} \sum_{\bar{\bm{s}}^2 \in \{0, 1\}^n} P\{\tilde{S} = \tilde{\bm{s}}^1\} P (\bm{s}, \bar{\bm{s}}^2),
\end{align*}
where the entries of $\bar{\bm{s}}^2$ are identical to the first $n$ entries of $\tilde{\bm{s}}^2$. Hence, 
\begin{align*}
&\quad~P\{\bar{S} = \bar{\bm{s}} | S = \bm{s}\} \\
&=\frac{P\{\tilde{S} = (\bar{\bm{s}}^T~\bm{s}^T)^T\}}{P\{S = \bm{s}\}}\\
&=\frac{\sum_{\tilde{\bm{s}} \in \mathscr{S}_1} P\{\tilde{S} = \tilde{\bm{s}}\} P (\bm{s}, \bar{\bm{s}})}{\sum_{\tilde{\bm{s}}^1 \in \mathscr{S}_1} \sum_{\bar{\bm{s}}^2 \in \{0, 1\}^n} P\{\tilde{S} = \tilde{\bm{s}}^1\} P (\bm{s}, \bar{\bm{s}}^2)}\\
&=\frac{P (\bm{s}, \bar{\bm{s}}) \sum_{\tilde{\bm{s}} \in \mathscr{S}_1} P\{\tilde{S} = \tilde{\bm{s}}\} }{\sum_{\tilde{\bm{s}}^1 \in \mathscr{S}_1} P\{\tilde{S} = \tilde{\bm{s}}^1\} \sum_{\bar{\bm{s}}^2 \in \{0, 1\}^n} P (\bm{s}, \bar{\bm{s}}^2)}\\
&=\frac{P (\bm{s}, \bar{\bm{s}}) \sum_{\tilde{\bm{s}} \in \mathscr{S}_1} P\{\tilde{S} = \tilde{\bm{s}}\} }{\sum_{\tilde{\bm{s}}^1 \in \mathscr{S}_1} P\{\tilde{S} = \tilde{\bm{s}}^1\} \cdot 1} = P (\bm{s}, \bar{\bm{s}}).
\end{align*}
\hfill$\Box$

\emph{\textbf{Proof of Theorem \ref{ident}}:}

From \eqref{gi} in the proof of Theorem \ref{MC}, we have the following 
\[
P\{S_1 = \bm{e}_i | S_0 = \bm{e}_j\} = (1 - \Phi(c_i - a_{ij})) \prod_{l \not = i} \Phi(c_l - a_{lj}),\]
\[P\{S_1 = \bm{0} | S_0 = \bm{e}_j\} = \prod_{1 \le l \le n} \Phi(c_l - a_{lj}), \]
\begin{align*}
P\{S_1 = \bm{e}_i | S_0 = \bm{e}_j + \bm{e}_k\} &= (1 - \Phi(c_i - a_{ij} - a_{ik})) \cdot \\
&\quad~ \prod_{l \not = i} \Phi(c_l - a_{lj} - a_{lk})\\
P\{S_1 = \bm{0} | S_0 = \bm{e}_j + \bm{e}_k\} &= \prod_{1 \le l \le n} \Phi(c_l - a_{lj} - a_{lk}),
\end{align*}
where $1 \le i, j, k \le n$ and $k \not = j$, and the same for $\{\hat{S}_t\}$. Here $\Phi$ is the c.d.f. of standard normal distribution.

Suppose that $\{S_t\}$ and $\{\hat{S}_t\}$ have the same probability transition matrices. From Assumption \ref{standardnormalasmp} and the above equations, it follows that
\begin{align*}
\Phi(c_i - a_{ij}) &= \Phi(\hat{c}_i - \hat{a}_{ij}) \\
\Phi(c_i - a_{ij} - a_{ik}) &= \Phi(\hat{c}_i - \hat{a}_{ij} - \hat{a}_{ik}),
\end{align*}
where $1 \le i, j, k \le n$ and $k \not = j$.
Hence by the strictly increasing property of $\Phi$, 
\begin{align*}
c_i - a_{ij} &= \hat{c}_i - \hat{a}_{ij} \\
c_i - a_{ij} - a_{ik} &= \hat{c}_i - \hat{a}_{ij} - \hat{a}_{ik},
\end{align*}
where $1 \le i, j, k \le n$ and $k \not = j$. Therefore, if we set $j = k + 1$ when $k < n$, and $j = 1$ when $k = n$, then we have for all $i, k \in \{1, \dots, n\}$, $a_{ik} = \hat{a}_{ik}$. Consequently $c_i = \hat{c}_i$ holds for all $i \in \{1, \dots, n\}$.
\hfill$\Box$

\bibliographystyle{ieeetr}
\bibliography{interpersonal}

\begin{thebibliography}{10}

\bibitem{d2000genetic}
P.~Dhaeseleer, S.~Liang, and R.~Somogyi, ``Genetic network inference: from
  co-expression clustering to reverse engineering,'' {\em Bioinformatics},
  vol.~16, no.~8, pp.~707--726, 2000.

\bibitem{ravazzi2017learning}
C.~Ravazzi, R.~Tempo, and F.~Dabbene, ``Learning influence structure in sparse
  social networks,'' {\em IEEE Transactions on Control of Network Systems},
  2017.

\bibitem{timme2014revealing}
M.~Timme and J.~Casadiego, ``Revealing networks from dynamics: an
  introduction,'' {\em Journal of Physics A: Mathematical and Theoretical},
  vol.~47, no.~34, p.~343001, 2014.

\bibitem{nozari2018network}
E.~Nozari, Y.~Zhao, and J.~Cort{\'e}s, ``Network identification with latent
  nodes via autoregressive models,'' {\em IEEE Transactions on Control of
  Network Systems}, vol.~5, no.~2, pp.~722--736, 2018.

\bibitem{sharf2018network}
M.~Sharf and D.~Zelazo, ``Network identification: A passivity and network
  optimization approach,'' in {\em 2018 IEEE Conference on Decision and Control
  (CDC)}, pp.~2107--2113, IEEE, 2018.

\bibitem{segarra2017network}
S.~Segarra, M.~T. Schaub, and A.~Jadbabaie, ``Network inference from consensus
  dynamics,'' in {\em 2017 IEEE 56th Annual Conference on Decision and Control
  (CDC)}, pp.~3212--3217, IEEE, 2017.

\bibitem{ravazzi2018randomized}
C.~Ravazzi, S.~Hojjatinia, C.~M. Lagoa, and F.~Dabbene, ``Randomized opinion
  dynamics over networks: influence estimation from partial observations,'' in
  {\em 2018 IEEE Conference on Decision and Control (CDC)}, pp.~2452--2457,
  IEEE, 2018.

\bibitem{dong2018identification}
Y.~Dong, W.~Zhao, {\em et~al.}, ``The identification of social networks by the
  least-square algorithm,'' in {\em 2018 37th Chinese Control Conference
  (CCC)}, pp.~1931--1936, IEEE, 2018.

\bibitem{frasca2009average}
P.~Frasca, R.~Carli, F.~Fagnani, and S.~Zampieri, ``Average consensus on
  networks with quantized communication,'' {\em International Journal of Robust
  and Nonlinear Control: IFAC-Affiliated Journal}, vol.~19, no.~16,
  pp.~1787--1816, 2009.

\bibitem{ceragioli2018consensus}
F.~Ceragioli and P.~Frasca, ``Consensus and disagreement: The role of quantized
  behaviors in opinion dynamics,'' {\em SIAM Journal on Control and
  Optimization}, vol.~56, no.~2, pp.~1058--1080, 2018.

\bibitem{wang2010system}
L.~Y. Wang, G.~G. Yin, J.-F. Zhang, and Y.~Zhao, {\em System identification
  with quantized observations}.
\newblock Springer, 2010.

\bibitem{guo2015asymptotically}
J.~Guo, L.~Y. Wang, G.~Yin, Y.~Zhao, and J.-F. Zhang, ``Asymptotically
  efficient identification of fir systems with quantized observations and
  general quantized inputs,'' {\em Automatica}, vol.~57, pp.~113--122, 2015.

\bibitem{guo2013recursive}
J.~Guo and Y.~Zhao, ``Recursive projection algorithm on fir system
  identification with binary-valued observations,'' {\em Automatica}, vol.~49,
  no.~11, pp.~3396--3401, 2013.

\bibitem{wang2018convergence}
T.~Wang, M.~Hu, and Y.~Zhao, ``Convergence properties of recursive projection
  algorithm for system identification with binary-valued observations,'' in
  {\em 2018 Chinese Automation Congress (CAC)}, pp.~2961--2966, IEEE, 2018.

\bibitem{marelli2013identification}
D.~Marelli, K.~You, and M.~Fu, ``Identification of arma models using
  intermittent and quantized output observations,'' {\em Automatica}, vol.~49,
  no.~2, pp.~360--369, 2013.

\bibitem{zhao2017recursive}
W.~Zhao, H.~Chen, R.~Tempo, and F.~Dabbene, ``Recursive nonparametric
  identification of nonlinear systems with adaptive binary sensors,'' {\em IEEE
  Transactions on Automatic Control}, vol.~62, no.~8, pp.~3959--3971, 2017.

\bibitem{godoy2011identification}
B.~I. Godoy, G.~C. Goodwin, J.~C. Ag{\"u}ero, D.~Marelli, and T.~Wigren, ``On
  identification of fir systems having quantized output data,'' {\em
  Automatica}, vol.~47, no.~9, pp.~1905--1915, 2011.

\bibitem{aguero2017based}
J.~C. Ag{\"u}ero, K.~Gonz{\'a}lez, and R.~Carvajal, ``Em-based identification
  of arx systems having quantized output data,'' {\em IFAC-PapersOnLine},
  vol.~50, no.~1, pp.~8367--8372, 2017.

\bibitem{song2018recursive}
Q.~Song, ``Recursive identification of systems with binary-valued outputs and
  with arma noises,'' {\em Automatica}, vol.~93, pp.~106--113, 2018.

\bibitem{pan2012modeling}
W.~Pan, W.~Dong, M.~Cebrian, T.~Kim, J.~H. Fowler, and A.~S. Pentland,
  ``Modeling dynamical influence in human interaction: Using data to make
  better inferences about influence within social systems,'' {\em IEEE Signal
  Processing Magazine}, vol.~29, no.~2, pp.~77--86, 2012.

\bibitem{chen2002stochastic}
H.-F. Chen, {\em Stochastic approximation and its applications}.
\newblock Kluwer, Boston, MA, 2002.

\bibitem{godoy2014identification}
B.~I. Godoy, J.~C. Ag{\"u}ero, R.~Carvajal, G.~C. Goodwin, and J.~I. Yuz,
  ``Identification of sparse fir systems using a general quantisation scheme,''
  {\em International Journal of Control}, vol.~87, no.~4, pp.~874--886, 2014.

\bibitem{burgoon2017social}
J.~K. Burgoon, N.~Magnenat-Thalmann, M.~Pantic, and A.~Vinciarelli, {\em Social
  Signal Processing}.
\newblock Cambridge University Press, 2017.

\bibitem{asavathiratham2001influence}
C.~Asavathiratham, {\em The influence model: A tractable representation for the
  dynamics of networked markov chains}.
\newblock PhD thesis, Massachusetts Institute of Technology, 2001.

\bibitem{meyn2012markov}
S.~P. Meyn and R.~L. Tweedie, {\em Markov chains and stochastic stability}.
\newblock Springer Science \& Business Media, 2012.

\bibitem{ljung1987system}
L.~Ljung, {\em System identification: theory for the user}.
\newblock Prentice-hall, 1987.

\bibitem{friedkin1990social}
N.~E. Friedkin and E.~C. Johnsen, ``Social influence and opinions,'' {\em
  Journal of Mathematical Sociology}, vol.~15, no.~3-4, pp.~193--206, 1990.

\bibitem{levin2017markov}
D.~A. Levin and Y.~Peres, {\em Markov chains and mixing times}, vol.~107.
\newblock American Mathematical Soc., 2017.

\bibitem{ferguson2017course}
T.~S. Ferguson, {\em A course in large sample theory}.
\newblock Routledge, 2017.

\bibitem{zhao2016iterative}
Y.~Zhao, W.~Bi, and T.~Wang, ``Iterative parameter estimate with batched
  binary-valued observations,'' {\em Science China Information Sciences},
  vol.~59, no.~5, p.~052201, 2016.

\bibitem{bertsekas2003convex}
D.~P. Bertsekas, A.~Nedi, A.~E. Ozdaglar, {\em et~al.}, ``Convex analysis and
  optimization,'' 2003.

\bibitem{malliavin2012integration}
P.~Malliavin, {\em Integration and probability}, vol.~157.
\newblock Springer Science \& Business Media, 2012.

\bibitem{blum1963strong}
J.~Blum, D.~L. Hanson, and L.~H. Koopmans, ``On the strong law of large numbers
  for a class of stochastic processes,'' {\em Zeitschrift f{\"u}r
  Wahrscheinlichkeitstheorie und verwandte Gebiete}, vol.~2, no.~1, pp.~1--11,
  1963.

\bibitem{doukhan1994mixing}
P.~Doukhan, {\em Mixing}.
\newblock Springer, 1994.

\end{thebibliography}

\end{document}